\documentclass[sigconf]{acmart}
\usepackage{hhline}
\usepackage{xcolor}
\usepackage{colortbl}
\usepackage{subfig}
\usepackage{comment}
\usepackage{enumitem}
\usepackage[ruled,vlined]{algorithm2e}
\usepackage{multirow}
\usepackage{graphicx}
\graphicspath{{./figures}}

\usepackage{tikz}

\newcommand{\pname}{\textit{gZCCL}}
\newcommand{\cname}{\textit{C-Coll}}

\newcommand{\ttf}[1]{$\mathtt{#1}$}

\AtBeginDocument{%
  }

\copyrightyear{2024}
\acmYear{2024}
\setcopyright{usgovmixed}
\acmConference[ICS '24]{Proceedings of the 38th ACM International Conference on Supercomputing}{June 4--7, 2024}{Kyoto, Japan}
\acmBooktitle{Proceedings of the 38th ACM International Conference on Supercomputing (ICS '24), June 4--7, 2024, Kyoto, Japan}\acmDOI{10.1145/3650200.3656636}
\acmISBN{979-8-4007-0610-3/24/06}

\begin{document}

\title{gZCCL: Compression-Accelerated Collective Communication Framework for GPU Clusters}

\author{Jiajun Huang}
\email{jhuan380@ucr.edu}
\orcid{0000-0001-5092-3987}
\affiliation{%
  \institution{University of California, Riverside}
  \city{Riverside}
  \country{United States of America}
}

\author{Sheng Di}
\email{sdi1@anl.gov}
\orcid{0000-0002-9935-5674}
\affiliation{%
  \institution{Argonne National Laboratory}
  \city{Lemont}
  \country{United States of America}
}

\author{Xiaodong Yu}
\email{xyu38@stevens.edu}
\orcid{0000-0001-6244-1264}
\affiliation{%
  \institution{Stevens Institute of Technology}
  \city{Hoboken}
  \country{United States of America}
}

\author{Yujia Zhai}
\email{yzhai015@ucr.edu}
\orcid{0000-0002-2688-8058}
\affiliation{%
  \institution{University of California, Riverside}
  \city{Riverside}
  \country{United States of America}
}

\author{Jinyang Liu}
\email{jliu447@ucr.edu}
\orcid{0000-0003-0177-502X}
\affiliation{%
  \institution{University of California, Riverside}
  \city{Riverside}
  \country{United States of America}
}

\author{Yafan Huang}
\email{yafan-huang@uiowa.edu}
\orcid{0000-0001-7370-6766}
\affiliation{%
  \institution{University of Iowa}
  \city{Iowa City}
  \country{United States of America}
}

\author{Ken Raffenetti}
\email{raffenet@anl.gov}
\orcid{0009-0003-4705-2713}
\affiliation{%
  \institution{Argonne National Laboratory}
  \city{Lemont}
  \country{United States of America}
}

\author{Hui Zhou}
\email{zhouh@anl.gov}
\orcid{0000-0002-4422-2911}
\affiliation{%
  \institution{Argonne National Laboratory}
  \city{Lemont}
  \country{United States of America}
}

\author{Kai Zhao}
\email{kzhao@cs.fsu.edu}
\orcid{0000-0001-5328-3962}
\affiliation{%
  \institution{Florida State University}
  \city{Tallahassee}
  \country{United States of America}
}

\author{Xiaoyi Lu}
\email{xiaoyi.lu@ucmerced.edu}
\orcid{0000-0001-7581-8905}
\affiliation{%
  \institution{University of California, Merced}
  \city{Merced}
  \country{United States of America}
}

\author{Zizhong Chen}
\email{chen@cs.ucr.edu}
\orcid{0000-0003-2578-4940}
\affiliation{%
  \institution{University of California, Riverside}
  \city{Riverside}
  \country{United States of America}
}

\author{Franck Cappello}
\email{cappello@mcs.anl.gov}
\orcid{0000-0002-7890-3934}
\affiliation{%
  \institution{Argonne National Laboratory}
  \city{Lemont}
  \country{United States of America}
}

\author{Yanfei Guo}
\email{yguo@anl.gov}
\orcid{0000-0002-3731-5423}
\affiliation{%
  \institution{Argonne National Laboratory}
  \city{Lemont}
  \country{United States of America}
}

\author{Rajeev Thakur}
\email{thakur@anl.gov}
\orcid{0000-0002-5532-3048}
\affiliation{%
  \institution{Argonne National Laboratory}
  \city{Lemont}
  \country{United States of America}
}
\renewcommand{\shortauthors}{Jiajun Huang et al.}

\begin{abstract}
GPU-aware collective communication has become a major bottleneck for modern computing platforms as GPU computing power rapidly rises. A traditional approach is to directly integrate lossy compression into GPU-aware collectives, which can lead to serious performance issues such as underutilized GPU devices and uncontrolled data distortion. In order to address these issues, in this paper, we propose \pname, a \textit{first-ever} general framework that designs and optimizes GPU-aware, compression-enabled collectives with an accuracy-aware design to control error propagation. To validate our framework, we evaluate the performance on up to 512 NVIDIA A100 GPUs with real-world applications and datasets. Experimental results demonstrate that our \pname-accelerated collectives, including both collective computation (Allreduce) and collective data movement (Scatter), can outperform NCCL as well as Cray MPI by up to 4.5$\times$ and 28.7$\times$, respectively. Furthermore, our accuracy evaluation with an image-stacking application confirms the high reconstructed data quality of our accuracy-aware framework.

\end{abstract}



\maketitle

\section{Introduction}
\label{sec:intro}

In the exascale computing era, efficient large-message collective communications are crucial for the performance of modern GPU-based supercomputers and clusters. This is particularly true for scientific applications and deep learning tasks that involve extensive data processing and exchange~\cite{awan2017s, abadi2016tensorflow, ayala2019impacts, jain2019scaling, abdelmoniem2021efficient}.
For example, the classic LSTM~\cite{Hochreiter1997LSTM} model used in the language modeling task can contain more than 66 million parameters and the communication overhead can be as high as 94\%~\cite{abdelmoniem2021efficient}, increasing the need for optimizing GPU-aware collective communication for large messages~\cite{chunduri2018characterization, Bayatpour2018SALaR}.

For GPU-aware collective communication, numerous researchers are actively working on mitigating network congestion in large-message collectives. Network saturation is often the major bottleneck because of limited network bandwidth. For example, even with advanced networks, such as HPE Slingshot 10, the network bandwidth is only about 100 Gbps~\cite{De_Sensi_2020Slingshot}. A straightforward solution is designing large-message collective communication algorithms that can minimize the transferred data volume instead of latency~\cite{Alm05BlueGene, thakur2005optimization, patarasuk2009bandwidth}. 
Another promising solution is shrinking the message size by error-bounded lossy compression techniques~\cite{Di2016SZ,Tao2017SZ,Zhao2020SZauto,Huang2023Exploring-Wavelet-Transform,Liu2024sigmod,Lindstrom2014ZFP}, as it can significantly reduce the data volume and maintain the data quality.

Previous lossy-compression-integrated approaches can be divided into two categories. The first is \textit{compression-enabled point-to-point communication} (namely CPRP2P) \cite{Zhou2021GPUCOMPRESSION}, which directly uses the 1D fixed-rate ZFP \cite{Lindstrom2014ZFP} to compress the data before it is sent and decompresses the received data after it is received. This method may cause significant overheads and unbounded errors in the collective communications as shown in~\cite{huang2023ccoll,Zhou2022GPUCOMPRESSIONALLTOALL}. The other category is to particularly optimize the \textit{compression-enabled collectives}. Zhou et al.\ \cite{Zhou2022GPUCOMPRESSIONALLTOALL} integrated the 1D fixed-rate ZFP \cite{Lindstrom2014ZFP} into MPI\_Alltoall on GPUs; however, this approach is limited to the Alltoall operation and CPU-centric staging algorithm and also results in the issue of unbounded error. Huang et al.~\cite{huang2023ccoll} designed an optimized general framework for compression-enabled collectives that can realize high performance for all MPI collectives with controlled errors. Nevertheless, this approach suffers from suboptimal performance on modern GPU clusters because of under-utilized GPU devices. 

Designing a GPU-aware compression-enabled collective communication system that realizes both high performance and controlled error propagation is non-trivial. There are three key challenges to address. 

\textbf{(1) How can we co-design and implement a compression-enabled collective algorithm that optimizes performance within modern GPU clusters?} For Allreduce operations, for example, state-of-the-art GPU-aware collective communication libraries, such as NCCL \cite{NCCL2023}
and MPICH \cite{MPICH2023}, adopt ring-based algorithms to optimize the transmission of large messages. However, it is unclear whether the ring-based model is the best fit when we include lossy compression techniques. In fact, unlike CPU, the GPU-based compression may easily face a low utilization issue, because of the inevitable GPU kernel-launch overhead and limited parallel design in GPU-based compression algorithms, which significantly lowers the performance. 

\textbf{(2) How can we optimize the redesigned algorithms to increase GPU utilization and decrease the required synchronizations and data transfers?} This is because unnecessary data transfers and synchronization can considerably increase the overall runtime and eliminate the opportunity for overlapping in the coordination of the host and device. 

\textbf{(3) How can we devise an accuracy-aware co-design that maintains data quality without sacrificing performance?} 
The accuracy of collective operations is at risk due to the data loss from GPU lossy compression. It is important to balance performance with accuracy.

To address the challenges mentioned above, this paper introduces a \textit{first-ever} generic high-performance framework, namely \pname, specifically designed for GPU-aware compression-accelerated collective communications.
Our contributions can be summarized in four key aspects:
\begin{itemize}
    \item To tackle challenge (1), we present two innovative algorithm design frameworks for classic collective operations, encompassing both collective computation and collective data movement. This proposal stems from a thorough analysis of the limitations in traditional large-message algorithms. This is fundamental to various co-designed compression-enabled collective algorithms, which can increase device utilization, decrease times of compression/decompression, and maximize performance.     
    \item To address challenge (2), we develop a series of optimization strategies to improve performance. Specifically, we improve the error-bounded lossy compressor (cuSZp~\cite{cuSZp2023}) and develop a multi-stream version to suit the context of the two collective performance optimization frameworks. 
    For the data movement framework, we overlap the compression/decompression, kernel launching, and data movement, respectively. For the collective computation framework, we enable possible overlapping between compression, decompression, and communication, which can further reduce the collective runtime.
    \item To address challenge (3), we design various strategies to considerably control the error accumulation in the \pname\ framework. We carefully design the \pname\ framework with the error-bounded lossy compressor that always causes an unknown compressed data size instead of the fixed-rate compressor that leads to a pre-known output data size to ensure a bounded error. We also decrease the number of compression operations on purpose, which can effectively decrease the number of stacked errors during the communication pattern.
    \item We integrate \pname\ framework into numerous collective operations, including Allgather, Reduce\_scatter, Allreduce, and Scatter, and meticulously evaluate their performance using different real-world scientific datasets. Experiments with up to 512 NVIDIA A100 GPUs reveal that other related works suffer from undesirable performance degradation in both Allreduce and Scatter due to significant compression overhead, inefficient GPU utilization, or larger data transfer volume. In contrast, our \pname-based Allreduce (referred to as gZ-Allreduce) outperforms the Allreduce in Cray MPI and NCCL by 20.2$\times$ and 4.5$\times$, respectively. Our \pname-based Scatter (gZ-Scatter) operates 28.7$\times$ faster than the MPI\_Scatter in Cray MPI.
    We also utilize a real-world use case (i.e., image stacking analysis) to validate the practical effectiveness of gZ-Allreduce. It demonstrates a 1.69$\times$ performance gain over NCCL, while still preserving a high level of data integrity.
\end{itemize}

The rest of the paper is organized as follows: we introduce background and related work in Section \ref{sec:background} and detail our design and optimization in Section \ref{sec-design_and_optimizations}. Evaluation results are presented in Section \ref{exp-setup-sec} followed by conclusion and future work in Section \ref{sec:conclusion}.


\section{Background and Related Work}
\label{sec:background}
Researchers have long been interested in utilizing compression to enhance MPI communication performance, based on the two communication categories -- point-to-point communication and collective communication. 

For the first category, a typical latest related work is utilizing 1D fixed-rate ZFP to boost MPI communications on GPU clusters \cite{Zhou2021GPUCOMPRESSION}. Their approach, however, focuses on enhancing MPI point-to-point communication performance, yielding suboptimal performance in collective scenarios. Furthermore, their solution could not provide a bounded error due to its fixed-rate design that fixes the compressed data size rather than ensuring accuracy. In contrast, our collective framework integrates error-bounded lossy compression, guaranteeing both high-quality compression and high collective performance. Hence, we regard this work as orthogonal to ours.

As for the second category, several existing studies explored how to optimize the MPI collective performance particularly, while they are limited to either CPU-centric communication (i.e., all the data are transferred through the CPU essentially) and/or have the uncontrolled error propagation. Zhou et al. proposed several optimized MPI collective operations \cite{Zhou2022GPUCOMPRESSIONALLTOALL, Zhou2023IPDPS, Zhou2022HiPC} using fixed-rate compression, which leads to inferior compression quality and unbounded error aggregation. On the contrary, our general framework provides a detailed guideline for designing and optimizing compression-accelerated collective algorithms, maximizing the performance of both collective computation and collective data movement while featuring well-controlled data distortion. Hence, we categorize these works as orthogonal works to ours. In addition, Huang et al. proposed an error-controlled compression-enabled framework that is capable of achieving a high performance across all MPI collectives \cite{huang2023ccoll}. Their method, however, fails to solve the inefficient GPU utilization, synchronization, and device-host data transfer issues, resulting in suboptimal performance on GPU clusters. In contrast, our GPU-centric framework is capable of fully utilizing the computational power of GPUs, significantly lowering the amounts of required compression, synchronization, and device-host data transfer, leading to a remarkable performance improvement. 

In the following text, we mainly focus on optimizing the performance of collective communications on GPU clusters by error-bounded lossy compression. This is because prior research \cite{huang2023ccoll} already demonstrated that the error-bounded lossy compression brings a limited and controllable impact on the final accuracy of collective communications by both theoretical and experimental analysis. 

\section{{\pname} Design and Optimization}
\label{sec-design_and_optimizations}

\begin{figure}[ht]
    \centering
 {\includegraphics[width=1\linewidth]{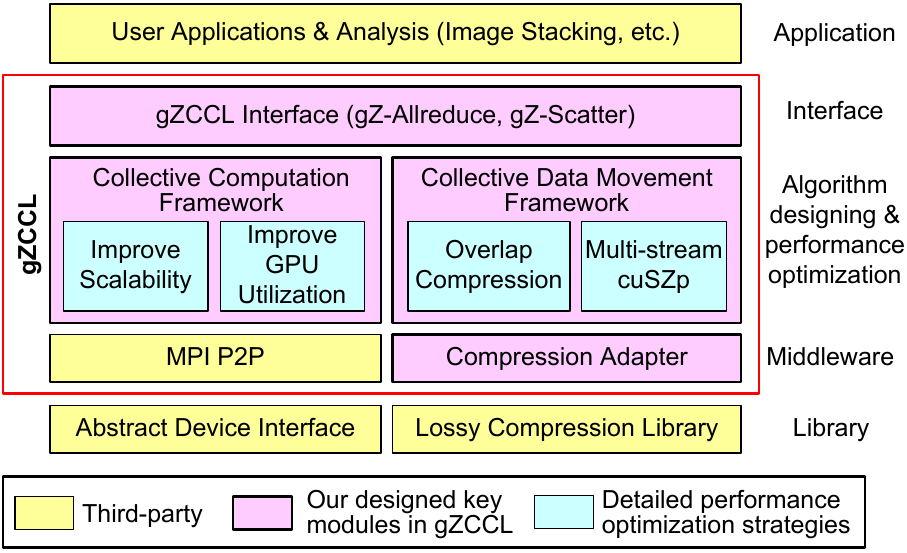}}
    \caption{\pname\ design architecture.} 
    \label{fig:architecture}
\end{figure}

In this section, we present our design and optimization strategies. Figure \ref{fig:architecture} shows the design architecture of \pname, where the newly designed modules are highlighted in purple boxes.
We develop an adapter that can run cuSZp \cite{cuSZp2023} more efficiently in regard to collective communications, to be detailed in Section~\ref{sec:SZp-adapter}.
We discuss our algorithm design as well as a series of performance optimization strategies, which are meticulously crafted for the two classic types of collectives -- collective computation and collective data movement. Details are described in Sections \ref{sec:gZCCL-algorithm-design} and \ref{sec:gZCCL-optimization}.

\subsection{\textbf{Analysis of existing compression-enabled GPU-aware collectives}}
In this section, we analyze the problems of prior solutions and provide a comprehensive performance breakdown to identify potential bottlenecks. 
\subsubsection{\textbf{Inefficient prior solutions in GPU-aware collectives}}

Lossy compression-enabled point-to-point communication (CPRP2\linebreak P) can decrease the transferred data volume \cite{Zhou2021GPUCOMPRESSION}, however, it faces huge accuracy loss and performance degradation in the collective scenario \cite{Zhou2022GPUCOMPRESSIONALLTOALL}. To solve these issues, \cname\ framework was proposed with two sub-frameworks: data movement framework and collective computation framework \cite{huang2023ccoll}. In the data movement framework, the data is pre-compressed and then sent along the communication patterns. Through this method, the huge compression overhead brought by the CPRP2P could be avoided. In the collective computation framework, the compression and communication costs are overlapped with each other, resulting in a better overall runtime. However, the direct implementation of the \cname\ framework may experience a huge performance degradation on modern GPU clusters due to two facts: 1. The current MPI collectives result in sub-optimal performance because all the temporary buffers are allocated on CPU, which means the data needs to be moved from GPU to CPU for the data to be transmitted over networks. Even though integrated compression can reduce the transferred message size, the device-host data movement cost can be significant. 2. The \cname\ framework does not address the inefficient GPU utilization problem and host-device synchronization issue, which may substantially degrade the collective performance. 


\subsubsection{\textbf{Identification of the bottlenecks in prior related works}}

\begin{figure}[ht] \centering
\subfloat[CPRP2P]
{
\includegraphics[width=0.25\textwidth]{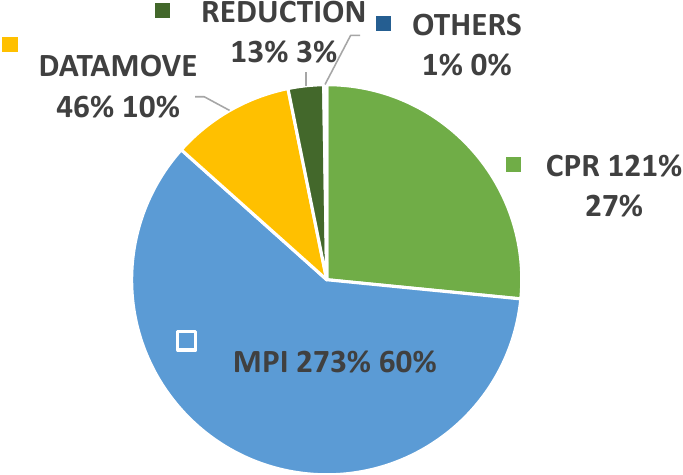}
}
\subfloat[{\cname}]
{
\includegraphics[width=0.24\textwidth]{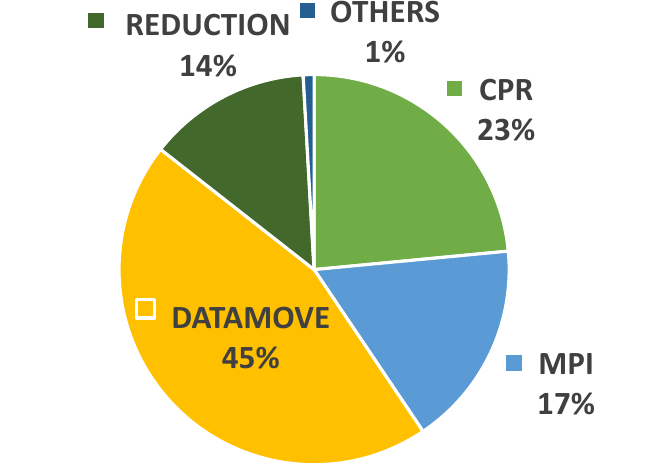}
}
\caption{Performance breakdown of Allreduce using CPRP2P and {\cname}: CPRP2P's first percentage is scaled to {\cname}'s runtime, and the second is scaled to its own.}
\label{fig:perf-breakdown}
\end{figure}

The ring-based Allreduce is a method commonly used in numerous state-of-the-art GPU-aware collective communication libraries such as NCCL \cite{NCCL2023} and MPICH \cite{MPICH2023}, particularly when optimizing large-message communications. This technique is composed of both data movement collective (Allgather) and collective computation (Reduce\_scatter), both of which have been optimized in \cname\ \cite{huang2023ccoll}.
Figure \ref{fig:perf-breakdown} presents a performance breakdown for the CPRP2P and {\cname} within the GPU-aware ring-based Allreduce algorithm. The evaluation is conducted utilizing 64 NVIDIA A100 GPUs, with 4 GPUs per node. When comparing CPRP2P versus {\cname}, it is evident that the latter significantly decreases the time cost in compression and decompression (CPR), resulting in overall enhanced performance.
However, it is notable that in {\cname}, the time required for host-device data transfer (DATAMOVE) is significant, accounting for nearly 45\% of the total runtime. In addition, the time consumed by compression and decompression (CPR) still remains substantial, occupying more than 23\% of the total time. This can be attributed to the inefficient utilization of GPUs. 
To rectify these problems, we present the {\pname} framework, whose design and implementation are detailed in subsequent sections.

\subsection{\textbf{Characterization of ring-based compression-enabled GPU-aware collectives}}
\subsubsection{\textbf{Traditional ring-based algorithms for long messages}}
Ring algorithms are widely acknowledged as the state-of-the-art solution for large-message collective communications such as Allgather, Reduce\_scatter, and Allreduce. In scenarios involving pure collective communications, ring approaches can significantly control the total data transfer volume, which can effectively control the network congestion when message sizes are large, thereby delivering optimal performance.
When CPU compression is employed, the CPU can be fully utilized for large message sizes, and the communication data volume can be substantially reduced, leading to a vast increase in overall collective performance. Prior research has shown that compression cost can be a dominant bottleneck in compression-enabled collectives. The reduction in communication volume in the ring-based algorithm design can lower the workload on the compressor, resulting in optimal performance. Hence, ring-based approaches are considered the most suitable algorithms for collectives integrated with CPU compression.

Taking into account modern GPUs \cite{Shanbhag2020Characteristics_of_CPUs_GPUs} features very high performance because of its performant single instruction, multiple threads (SIMT) architecture, adopting GPU-based lossy compression may further reduce the compression overhead intuitively, however, a key question arises: Can GPUs still be fully utilized in the compression-enabled ring-based algorithm? 
In fact, unlike CPUs which are often saturated, GPU performance is heavily dependent on the utilization rate. That is, a low utilization rate on GPU will increase the compression cost and lead to sub-optimal collective performance. To answer the above question, we need to characterize the performance of the lossy compressor.

\subsubsection{\textbf{Characterization of GPU lossy compressor}}
\label{sec:Characterization-SZp}
\begin{figure}[ht]
    \centering
    \hspace*{-4mm}
    {\includegraphics[scale=0.38]{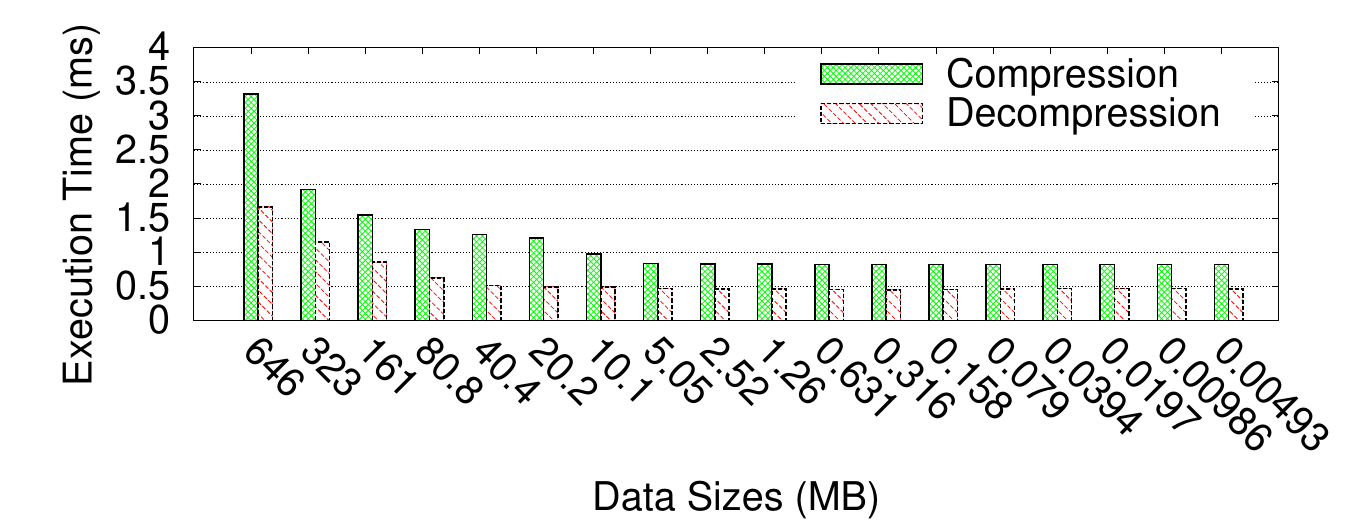}}
    \caption{Characterization of cuSZp compression and decompression execution time with uniform data.} 
    \label{fig-cuszp_speed}
\end{figure}

In this section, we detail the characterization of the GPU lossy compressor -- cuSZp \cite{cuSZp2023}, and this process is also applicable to other GPU compressors. Utilizing 646MB (the data size of the largest scientific dataset we use later) of synthetic data where all data points are uniformly distributed, we characterize the performance of cuSZp on an NVIDIA A100 GPU as shown in Figure \ref{fig-cuszp_speed}.
We observe that as data size decreases, execution time decreases for both compression and decompression kernels with a declining rate, and even stagnates when the data size is smaller than 5MB. This indicates that the GPU is not fully utilized, especially when the input data size is relatively small, and the utilization rate continues to drop with a decrease in message size.
However, an input message larger than 1MB is already considered a large message in collective communications, and the actual message to be sent/received or compressed during ring-like communication patterns is much smaller than the input message. This is because the original data is divided into small blocks for communications. Consequently, ring-based algorithms may result in relatively low GPU utilization, and we provide a more detailed discussion in the following text.

\subsubsection{\textbf{Ring-based collective computation}}
\label{sec:Ring-based-reduce-scatter}
In this section, we explore the limitations of the ring-based algorithms integrated with GPU compression, using the ring-based Reduce\_scatter operation as an illustrative example.
In the ring-based Reduce\_scatter operation, the input data, denoted by size $D$, is divided into $N$ small chunks, with $N$ being the process count. Each of these chunks undergoes a ring-like communication pattern for reduction across $N$$-$1 rounds. When the GPU compression is incorporated, each round provides a data chunk of size $D/N$ to the compression kernel, while an equal-sized output is produced by the decompression kernel. This mechanism necessitates a total of $N$$-$1 rounds of both compression and decompression.
Consequently, even when dealing with large message sizes like 646MB, the GPU experiences significantly poor utilization when the process count reaches approximately 128 ($646/5.05\approx128$), according to our previous analysis in Section \ref{sec:Characterization-SZp}. This results in compromised scalability. Further exacerbating this issue is the fact that the total number of decompression and compression operations is $N$$-$1, which scales linearly with the process count $N$.
Notably, this problem is not exclusive to the ring-based Reduce\_scatter operation. The widely-used ring-based Allreduce operation, which is composed of ring-based Allgather and Reduce\_scatter, is also plagued by these scalability and performance shortcomings. Therefore, the direct application of ring-based algorithms for collective computation with GPU compression may not always yield optimal results. It is hence vital to explore other algorithms that may offer superior performance.

\subsection{\textbf{Proposing the novel \pname\ framework}}
In this section, we delve into the details of our \pname\ framework. Our primary goal is to address and overcome the performance issues noted in the previous GPU-aware MPI collective framework that incorporates compression, such that a superior performance can be reached.

\subsubsection{\textbf{Getting rid of the traditional host-centric design}}
To circumvent the high cost of device-to-host data transfer inherent in traditional CPU-centric designs, we implement a GPU-centric design. Specifically, when GPU support is enabled, a sufficiently large GPU buffer pool is pre-allocated during the \ttf{MPI\_Init} function call. The size of this GPU buffer pool can be adjusted based on user input. Hence, GPU-aware MPI collectives can leverage these pre-allocated device buffers directly during function calls, rather than repeatedly allocating them amidst intensive communications. This is not only resource-intensive but also causes undesired host-device synchronization.
Moreover, current MPI implementations tend to use the host for carrying out reduction operations in collective computations. In response to this, we designed and implemented a GPU reduction kernel capable of processing data entirely on the device. With these optimizations, we successfully transition from the original host-centric algorithms and elevate the compression-enabled collectives to the device-centric level.

\subsubsection{\textbf{Adapting lossy compression to achieve high collective performance}}
\label{sec:SZp-adapter}
To improve collective performance in compression-enabled collectives, it is critical to adapt the lossy compression to suit the requirements of collective communications. We illustrate our customization and optimization strategies based on cuSZp, and the improvement strategies can also be applied to other lossy compressors. 

In the following, we analyze the potential performance issue of cuSZp, and then describe our improvement strategies.
In the cuSZp function \ttf{cuSZp\_compress\_deviceptr}, an initial step involves the allocation of a unified memory buffer known as \ttf{d\_cmpOffset}, accessible from both the device and host. This joint accessibility incurs implicit host-device data transfer, leading to suboptimal performance. To counteract this issue, we redesign cuSZp's data allocation process, liberating cuSZp from the constraints of unified memory. This modification results in a reduction of necessary data transfers, subsequently improving performance.
Moreover, cuSZp allocates temporary buffers to store compression-related parameters upon any invocation of the \ttf{cuSZp\_compress\_deviceptr} function. This procedure may block the host and also generates unwanted device overheads in collective scenarios where compression is frequently executed. To address this issue, our solution allocates a temporary buffer, which will be cleared and reused for any compression operations, so that the memory allocation costs can be reduced significantly also with data integrity. 

\subsubsection{\textbf{Two algorithm design frameworks}}
\label{sec:gZCCL-algorithm-design}
In this section, we describe the algorithm design inherent to our \pname.

\vspace{1mm}
\noindent\textbf{\textit{Exploring new metrics regarding GPU compression-enabled collective performance.}} As for the GPU compression-enable collective algorithms, there are several important new metrics that need to be addressed in particular. 

\textit{\textbf{Total compression cost}}. The compression cost is determined by two critical factors: per-compression time cost and the number of compression executions. As for the per-compression cost on GPU, it may face a low utilization issue when the input data is not large enough, as discussed in Section \ref{sec:Characterization-SZp}. For example, $10$ times of compression of $1$ MB data can be much more expensive than $1$ compression of $100$ MB data as shown in Figure \ref{fig-cuszp_speed}. As such, we should pay much attention to the number of times the data need to be compressed, in order to minimize the total compression cost. As verified in Section \ref{sec:Ring-based-reduce-scatter}, we demonstrate that large-message algorithms such as ring-based algorithms may result in low scalability with compression in some cases, which is due to the fact that they can result in more compression operations each with low GPU utilization. In the compression-enabled collectives, how often the compression is executed is closely related to the times of the data communications, which are generally optimized by the small-message algorithms. Thus, the conclusion is that, with GPU compression integrated, the small-message algorithms may outperform the large-message algorithms.

\textit{\textbf{Accuracy loss}}. Apart from the compression-related overheads, another concern of integrating lossy compression in the collectives is the accuracy loss caused by accumulated errors along with the intensive communications. Again, the large-message algorithms like the ring-based approach can introduce larger errors compared with the small-message algorithms such as the one based on the recursive-doubling algorithm, further degrading the reconstructed data quality. This is due to the fact that the ring-based algorithm requires $N$$-$1 times of compression/decompression and the recursive-doubling-based algorithm only needs $log N$ compression/decompression operations. Fortunately, the increased times of compression/decompression may not bring a huge accuracy difference statistically because the mathematical expectation of all accumulated errors is 0. Thus, we can achieve a high reconstructed data quality with the integration of lossy compression in the collective communications, which will be demonstrated later in Section \ref{sec:image_stacking}.

\vspace{1mm}
\noindent\textbf{\textit{Collective computation algorithm design framework.}} In the following discussion, we will employ the typical Allreduce operation as a case study to describe the algorithm design of our \pname\ framework in collective computation scenarios. In general, the recursive doubling algorithm is employed for short messages due to its optimized latency, whereas the previously-mentioned ring-based algorithm is used for large messages in Allreduce because of its ability to control the data transfer volume \cite{thakur2005optimization}.
The ring-based Allreduce operation consists of a Reduce\_scatter stage and an Allgather stage. In the Reduce\_scatter stage, $N$$-$1 compression/decompression operations are required, while the Allgather stage necessitates one compression and $N$$-$1 decompression operations \cite{huang2023ccoll}.
When compared with the $N$ compression operations and $N$$-$1 decompression operations required by the ring-based Allreduce algorithm, the recursive doubling algorithm involves only $log\ N$ communication steps or compression/decompression operations, where $N$ is the process count. As such, the recursive doubling algorithm exhibits superior scalability in terms of compression cost, especially when $D/N < 5MB$, where $D$ denotes the input data size.
However, when compressing the data with the data size being $D/N$ and the GPU utilization is high, the ring-based algorithm still outperforms the recursive doubling one as it can minimize both compression and communication workloads. In conclusion, the recursive doubling-based Allreduce algorithm delivers high scalability, while the ring-based one projects a high performance when GPU utilization is high.

\vspace{1mm}
\noindent\textbf{\textit{Collective data movement algorithm design framework.}} In this section, we delve into the algorithm design of our \pname\ framework in collective data movement scenarios. Generally, there are three types of collective data movement: one-to-all, all-to-one, and all-to-all. The all-to-all communication pattern is the most complex as it encapsulates both one-to-all and all-to-one communications. Accordingly, we select the extensively-used all-to-all communication operation -- Allgather -- as a case study to demonstrate the algorithm selection process in \pname. Note that this design can also be applied to other collectives.
In essence, the Bruck algorithm and the recursive doubling algorithm are optimized toward lowering latency, while the ring-based algorithm prioritizes minimizing data transfer volume. Unlike collective computation scenarios, data compression only happens at the beginning and the end of the collective data movement. For instance, the data in the Allgather operation of each process should be compressed first, then the compressed data is communicated between processes. After all communications are completed, each process decompresses the gathered compressed data to retrieve the original data.

In what follows, we extensively analyze which compression-enabled algorithm is the best fit for the Allgather operation.
Although the ring-based Allgather requires $N$$-$1 communication steps to finish, it only necessitates one compression and $N$$-$1 decompression operations. In addition, the $N$$-$1 decompression operations can be overlapped using multi-stream techniques to improve GPU utilization, which will be detailed in Section \ref{sec:data-movement-optimization}. In conclusion, the ring-based Allgather only suffers from inefficient GPU utilization in one compression operation, and it benefits from optimized data transfer volume.
Therefore, although the Bruck and recursive doubling algorithms exhibit the least communication steps or compression operations, they cannot further improve scalability and suffer from sub-optimal data transfer volume compared to the ring-based algorithm. As a result, the ring-based approach emerges as the optimal choice for the compression-integrated Allgather operation.

\subsubsection{\textbf{Two performance optimization frameworks}}
\label{sec:gZCCL-optimization}
\begin{figure}[ht]
    \centering
    {\includegraphics[width=1\linewidth]{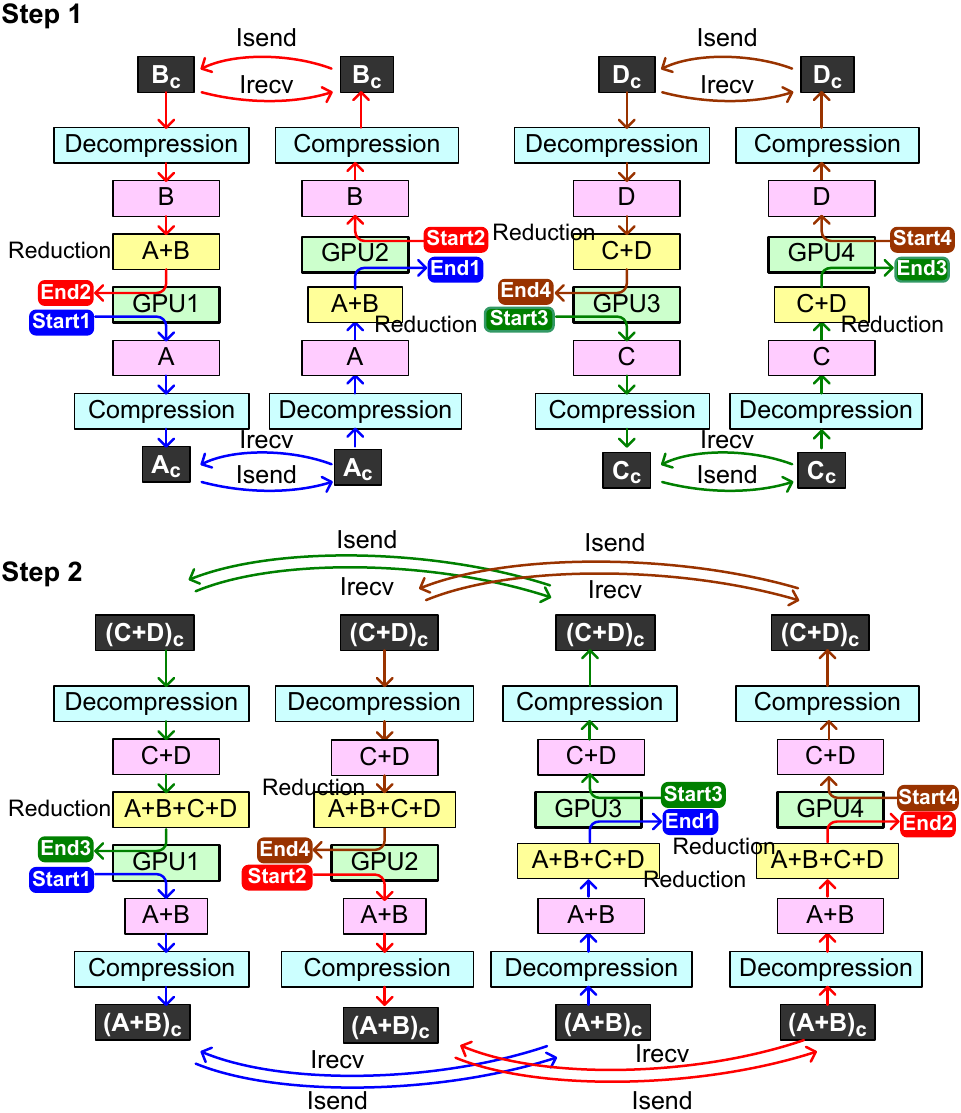}}
    \caption{Design of our \pname\ collective computation framework on compression-accelerated gZ-Allreduce. This example uses four GPUs/Processes.} 
    \label{fig-Recur-Doubling-Z-allreduce}
\end{figure}

In this section, we give a comprehensive discussion of the intricate optimization techniques that are integral to our \pname\ framework. By unveiling the technical underpinnings of our framework, we aim to provide an in-depth understanding of how our methods contribute to improved performance and efficiency in GPU-based computational systems.
\vspace{1mm}
\\
\textbf{\textit{Developing multi-stream lossy compression.}}
To facilitate multi-stream compression and decompression within collective communication, we need to tailor the lossy compressor, which originally operates using a single default GPU stream. For illustrative purposes, we mainly describe the compression procedure based on the state-of-the-art GPU-based compressor -- cuSZp as an example.  
We begin by delving into the source code to modify the cuSZp compression process, enabling it to accept a user-defined stream rather than operating exclusively on the default stream. This new stream-supported compression API is henceforth referred to as \ttf{cuSZp\_compress\_stream}.
To effectively overlap compression across different streams, it is imperative to ensure the absence of data races and undesired conflicts. Accordingly, we conduct a meticulous analysis and testing of the critical paths and data dependencies within cuSZp.
During this investigation, we find that beyond the standard \ttf{d\_oriData} (buffer of original data) and \ttf{d\_cmpBytes} (buffer of compressed data), cuSZp requires several distinct device buffers to store temporary information, including offsets of various compression blocks and flag bits. Consequently, we independently allocate buffers for each stream to avoid data conflicts in the multi-stream scenario. The decompression as well as other lossy compressors can be multi-streamed similarly, and we omit details due to space limit.
\vspace{1mm}
\\
\textbf{\textit{Collective computation performance optimization framework.}} 
In this section, we illustrate the \pname\ optimizations in the collective computation routines using the recursive doubling-based Allreduce as an example. Similar optimizations can be applied to other collective computation algorithms such as Reduce\_scatter. 
Figure \ref{fig-Recur-Doubling-Z-allreduce} illustrates the \pname\ implementation on the recursive doubling-based compression-enabled Allreduce operation (we call it gZ-Allreduce (ReDoub)). We first create one non-default stream and a set of temporary device buffers then reuse these GPU buffers for all the compression and decompression to avoid extra overheads. Then, the design contains two main stages, which will be described in the following text, where $N$ is the number of processes, and $r$ refers to the remainder of the process count taking away the maximum power of two: i.e., $r$ = $\min (N-2^k)$, where $k$ $\in$ $\mathbb{Z}_{+}$ and $k\leq \log_2 N$. 

In the first stage, we mainly handle the remainder processes ($r$ processes). In the case where the number of processes is not a power of two, all even-numbered processes with a rank (denoted $i$) lower than 2$r$ first asynchronously clear the temporary GPU buffers and launch the compression kernel on the non-default stream to compress their whole data and sending their compressed data to the process of rank $i$+1. Meanwhile, the odd-numbered processes pose non-blocking receive operations to obtain the compressed data and clear the GPU buffers for decompressing them on the non-default stream. Then, these even-numbered processes are suspended until the final stages, and the odd-numbered processes half their ranks ($i$=$i/2$). 

In the second stage, we handle the remaining power of two processes (i.e., $2^k$). For the processes with ranks $i$$\geq$2$r$, we update the ranks by $i$=$i$$-$$r$. Then, in each recursive doubling communication step, each process asynchronously memsets the temporary device buffers and launches the compression kernel on the non-default stream to compress the data. The compressed data is sent through a non-blocking send operation and another non-blocking receive operation is posed to receive the compressed data from another process. Upon the receiving of data, a clear operation and decompression kernel are launched to obtain the original data. Thereafter, the reduction kernel is launched on the non-default stream to reduce the decompressed data and data in the receive buffer. Unlike the ring-based case, each communication step requires sending/receiving the whole data instead of the divided data blocks, ensuring high GPU utilization. 

\vspace{1mm}
\noindent\textbf{\textit{Collective data movement performance optimization framework.}} 
\label{sec:data-movement-optimization}
In this section, we describe how we optimize collective data movements to enhance GPU utilization. We use the binomial tree-based \pname-accelerated Scatter/Scatterv as an example. Similar optimization can be applied to other collective data movement operations, such as Allgather.

\begin{figure}[ht]
    \centering
    {\includegraphics[width=0.9\linewidth]{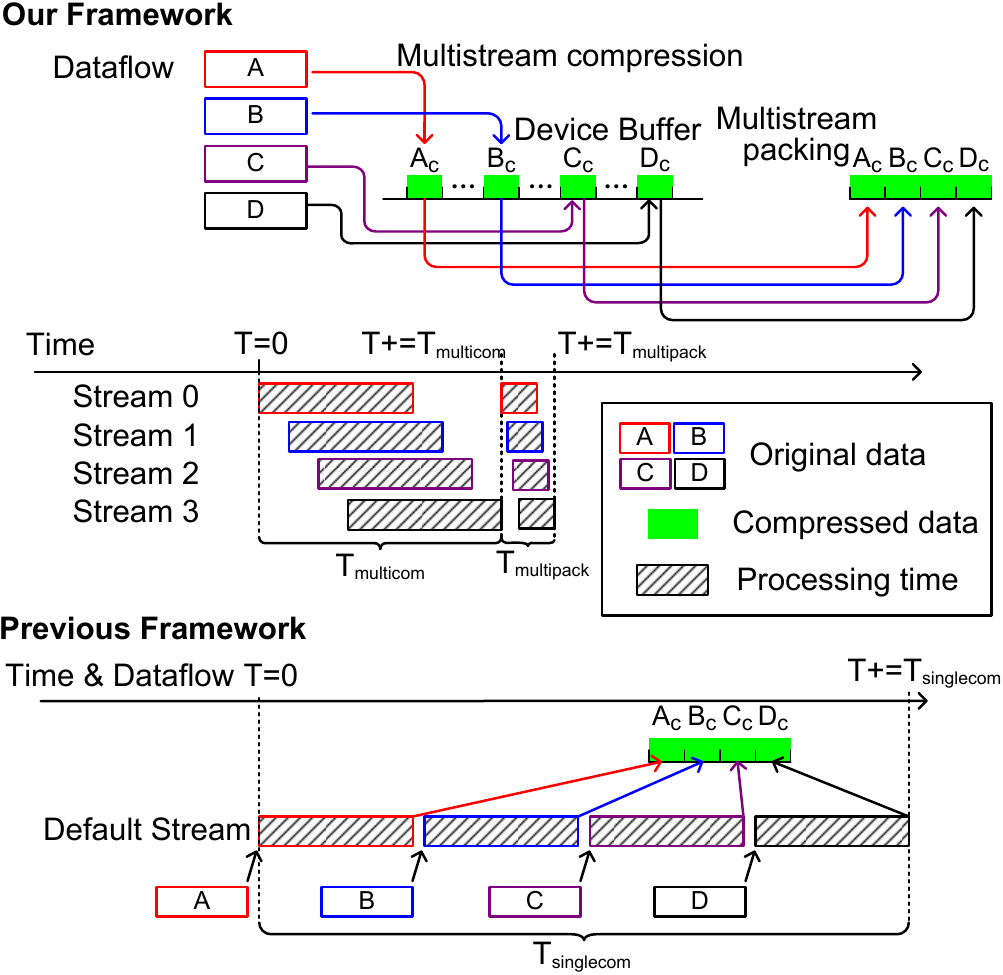}}
    \caption{Design of our \pname\ data movement framework on compression-accelerated gZ-Scatter. This example uses four GPUs/Processes.} 
    \label{fig-gZ-Scatter-design}
\end{figure}

We design our gZ-Scatter based on the binomial tree-based Scatter algorithm that is utilized in both short and long messages\cite{thakur2005optimization}. In Figure \ref{fig-gZ-Scatter-design}, we present the overall design of our gZ-Scatter. In this algorithm, the original data on the root process is distributed to other processes in a binomial tree communication pattern. An intuitive solution is compressing the original data as a whole and sending the compressed data by blocks to other processes, which however introduces two challenges. On the one hand, the compressed bytes contain some metadata that are essential for decompression. If the compressed data are directly divided into smaller blocks, the vital information will be lost. On the other hand, the original data distribution might not be uniform and the compressed data sizes for each block are not equal. As a result, it is impossible for us to correctly separate the compressed data into data blocks in this case. Thus, we need to individually compress the corresponding data blocks and then distribute them. 

To better explain our optimization and design, more details are shown as follows.
First, we create helper arrays on the CPU to store the compressed data sizes and the related global offsets on each process. Then, in each process, we create a stream array of size $N$, where $N$ is the size of the communicator. Additionally, we allocate two device buffer pointer arrays, also of size $N$, to store the offsets of compressed bytes and flag information for differing streams, respectively. In the root process, we launch the multi-stream compression kernel utilizing the independent device buffers and streams from $0$ to $N$$-$1 in the stream array. The compressed data for each stream is stored in the same device buffer based on the designated offset so that there are no data races. Then, we synchronize these streams with the host to make sure the multi-stream compression has finished. After that, we obtain the compressed data sizes and offsets of different streams and synchronize the information with other non-root processes. Then, we use asynchronous memcpys with different streams to pack these compressed data based on the compressed data offsets into another device buffer, so that they can be sent out in a continuous format. Finally, the data is distributed in a binomial tree communication pattern and the non-root processes utilize a non-default stream to decompress its own part of compressed data. In a nutshell, we have optimized the compression-enabled Scatter algorithm with overlapped compression, kernel launching, and data movements, resulting in improved performance.

\section{Experimental Evaluation}\label{exp-setup-sec}

We present and discuss the evaluation results as follows.

\subsection{Experimental Setup}
\label{sec:experimental-setup}
We perform the evaluation on a GPU supercomputer that involves 512 NVIDIA A100 80G GPUs (128 nodes each with 4 GPUs, specifically), which features both internode communication and intranode communication. These computational nodes are interconnected via the HPE Slingshot 10 interconnect, providing a network bandwidth of 100 Gbps. Unless specified, the absolute error bound of compression is set to 1E-4, because the image reconstruction quality is already superior with 2E-4 error bound, which will be demonstrated later in Figure \ref{fig-image-stacking}.
Two distinct RTM datasets \cite{Kayum2020RTM}, originating from the real-world 3D SEG/EAGE Overthrust model, are generated under two different simulation settings. Table \ref{tab:compression-ratio-quality} exhibits the average compression ratio and PSNR that cuSZp can achieve for these datasets, where ABS denotes the absolute error bound.

\begin{table}[ht]
\centering
\caption{Compression ratio (CPR) and quality (PSNR).}
\label{tab:compression-ratio-quality}
\resizebox{0.95\columnwidth}{!}{%
\begin{tabular}{|c|cc|cc|}
\hline
 & \multicolumn{2}{c|}{\textbf{Simulation Setting 1}} & \multicolumn{2}{c|}{\textbf{Simulation Setting 2}} \\ \hline
\textbf{Dimensions} & \multicolumn{2}{c|}{\textbf{449X449X235}} & \multicolumn{2}{c|}{\textbf{849X849X235}} \\ \hline\hline
\textbf{ABS} & \multicolumn{1}{c|}{\textbf{CPR}} & \textbf{PSNR} & \multicolumn{1}{c|}{\textbf{CPR}} & \textbf{PSNR} \\ \hline
1E-3 & \multicolumn{1}{c|}{92.28} & 53.23 & \multicolumn{1}{c|}{94.41} & 53.41 \\ \hline
1E-4 & \multicolumn{1}{c|}{73.35} & 65.67 & \multicolumn{1}{c|}{63.94} & 70.38 \\ \hline
1E-5 & \multicolumn{1}{c|}{55.65} & 78.83 & \multicolumn{1}{c|}{46.74} & 88.57 \\ \hline
\end{tabular}%
}
\end{table}

\subsection{Evaluating the GPU-centric design}

\begin{figure}[ht]
    \centering
    \hspace{-10mm}
    \subfloat[Simulation Setting 1]{
        \includegraphics[scale=0.38]{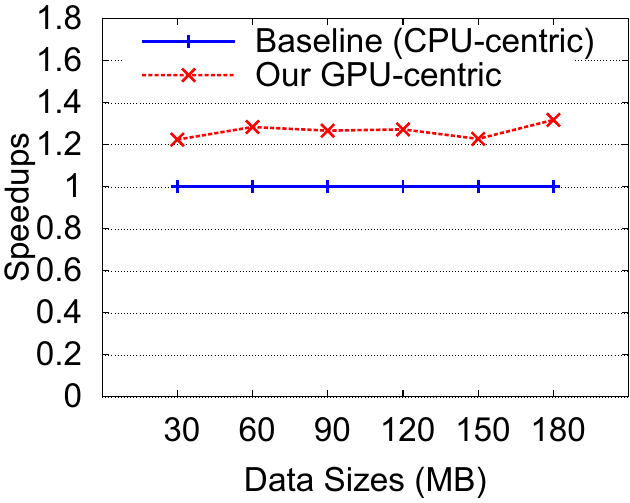}
        \label{fig-GPU-centric-small}
    }
    \subfloat[Simulation Setting 2]{
        \includegraphics[scale=0.395]{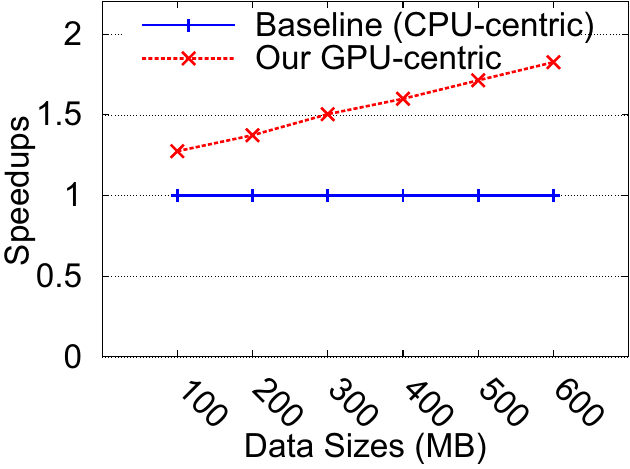}
        \label{fig-GPU-centric}
    }
    \hspace{-8mm}    
    \caption{Performance evaluation of our GPU-centric design using two different scientific datasets.}
    \label{fig-GPU-centric-performance}
\end{figure}

First of all, we present the performance evaluation of our proposed GPU-centric design compared with the traditional CPU-centric solution on 64 NVIDIA A100 GPUs across 16 nodes, using two different scientific datasets and the Allreduce collective operation. As mentioned in Section \ref{sec:intro} and Section \ref{sec:background}, many of the existing related works \cite{Zhou2022GPUCOMPRESSIONALLTOALL, huang2023ccoll} are dependent on the CPU-centric communication design. As shown in Figure \ref{fig-GPU-centric}, it is noticeable that the speedups of GPU-centric design over the CPU-centric solution increase with the expansion of the data sizes, culminating in a 1.82$\times$ performance improvement for the data size of 600 MB. This trend is also observed in Figure \ref{fig-GPU-centric-small}, where the speedup can reach up to 1.32$\times$ with the largest 180 MB data size. As data size increases, the demand of  intensive host-device data movement escalates in the CPU-centric design, which may cause an increasing PCIe congestion and reduction cost. This creates a pronounced bottleneck for the overall collective performance. 
To mitigate the substantial cost of host-device data transfer, our GPU-centric design does not depend on CPU-based communication, totally eliminating the data movement cost between CPU and GPU. Moreover, our design can significantly mitigate reduction operation cost, further boosting the speedup, especially with the growth of data size.  

\subsection{Evaluating the optimized redesigned GPU compression-enabled collective algorithms}
We evaluate the performance of our optimized, compression-integra\linebreak ted collective algorithms using 64 NVIDIA A100 GPUs.
\subsubsection{Collective computation}

\begin{figure}[ht]
    \centering
    \hspace{-10mm}
    \subfloat[Simulation Setting 1]{
        \includegraphics[scale=0.38]{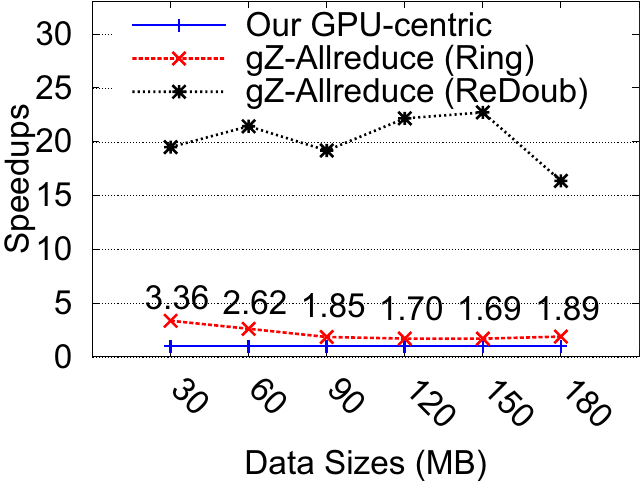}
        \label{fig-gZ-Allreduce-small}
    }
    \subfloat[Simulation Setting 2]{
        \includegraphics[scale=0.38]{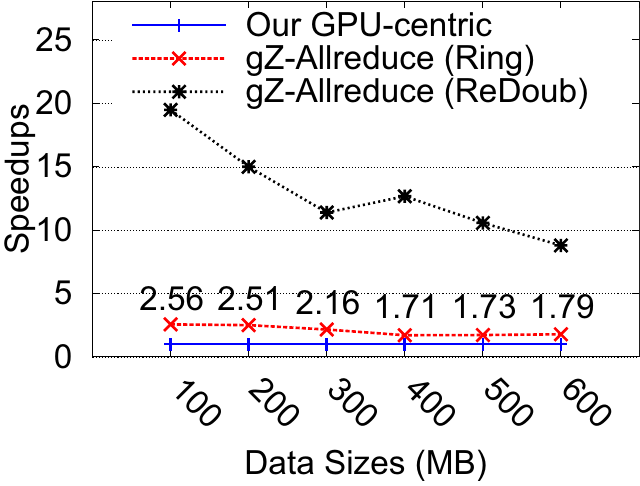}
        \label{fig-gZ-Allreduce}
    }
    \hspace{-8mm}
    \caption{Performance evaluation of our \pname\ collective computation framework using Allreduce operation.}
    \label{fig-gZ-Allreduce-performance}
\end{figure}

In this section, we evaluate our optimized redesigned compression-enabled collective computation algorithms using the widely-used Allreduce operation. 
Both Figure \ref{fig-gZ-Allreduce-small} and \ref{fig-gZ-Allreduce} reveal that our optimized solution -- gZ-Allreduce (Ring) surpasses our original GPU-centric approach by up to 3.36$\times$. This is because our solution improves GPU utilization. Specifically, we overlap the decompression and kernel launching in the Allgather stage and facilitate potential overlapping among compression, decompression, and communication in the Reduce\_scatter stage. Furthermore, the newly designed gZ-Allreduce (ReDoub) achieves an even higher performance enhancement compared to gZ-Allreduce (Ring), attaining up to 22.7$\times$ speedup compared to our original GPU-centric approach. We explain the reasons as follows. To tackle the inefficient device utilization in ring-based Allreduce, we design and optimize a novel recursive doubling-based compression-enabled algorithm, with the aim of improving scalability, maximizing performance, and preserving accuracy. 
However, it is worth noting that the speedup of both gZ-Allreduce methods generally decreases as the data size increases. This is because the problem of inefficient GPU utilization can be mitigated by larger message sizes, and the performance improvement resulting from higher GPU utilization would consequently decrease.

\subsubsection{Collective data movement}

\begin{figure}[ht]
    \centering
    \hspace{-10mm}
    \subfloat[Simulation Setting 1]{
        \includegraphics[scale=0.38]{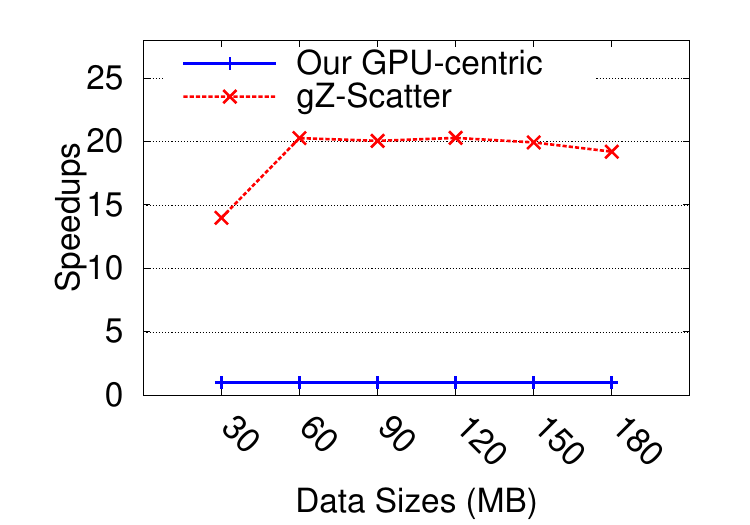}
        \label{fig-gZ-Scatter-small}
    }
    \hspace{-8mm}
    \subfloat[Simulation Setting 2]{
        \includegraphics[scale=0.38]{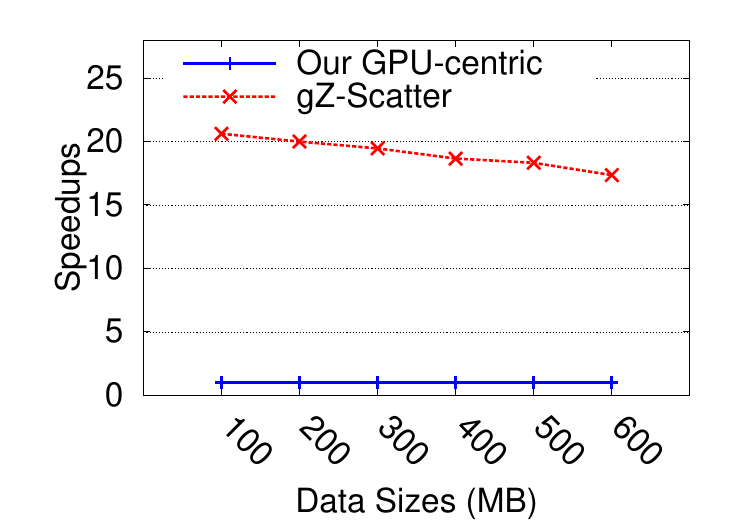}
        \label{fig-gZ-Scatter}
    }
    \hspace{-8mm}
    \caption{Performance evaluation of our \pname\ collective data movement framework using Scatter operation.}
    \label{fig-gZ-Scatter-performance}
\end{figure}

In this section, the performance of our optimized, redesigned compression-integrated collective data movement algorithms is demonstrated, using the classic Scatter operation. 
From Figure \ref{fig-gZ-Scatter-small} and Figure \ref{fig-gZ-Scatter}, we notice that gZ-Scatter exhibits substantial speedups in both datasets, obtaining up to 20.3$\times$ and 20.6$\times$ improved performance in the two simulation settings, respectively. This is because, in the gZ-Scatter algorithm, we overlap compression, kernel launching, and data movements, leading to enhanced device utilization, diminished host-device synchronization, and reduced device-device data movement cost. Similar to the collective computation scenario, with increasing data sizes, the performance boost slightly diminishes, with a minimum of 17.4$\times$ at 600 MB as depicted in Figure \ref{fig-gZ-Scatter}. This reason is that a larger input data size can better saturate the device, thereby mitigating the performance enhancement derived from our \pname\ design. 

\subsection{Comparisons of \pname\ with other collective communication libraries}
In this section, we compare the performance of our \pname\ framework with other state-of-the-art GPU communication libraries, such as the widely-utilized NCCL and CUDA-aware Cray MPI.
\subsubsection{Collective computation}
In this section, the performance of our \pname\ collective computation framework is compared with both NCCL and Cray MPI, using the prevalent Allreduce operation. 

\begin{figure}[ht]
    \centering
    \hspace*{-5mm}
    {\includegraphics[scale=0.38]{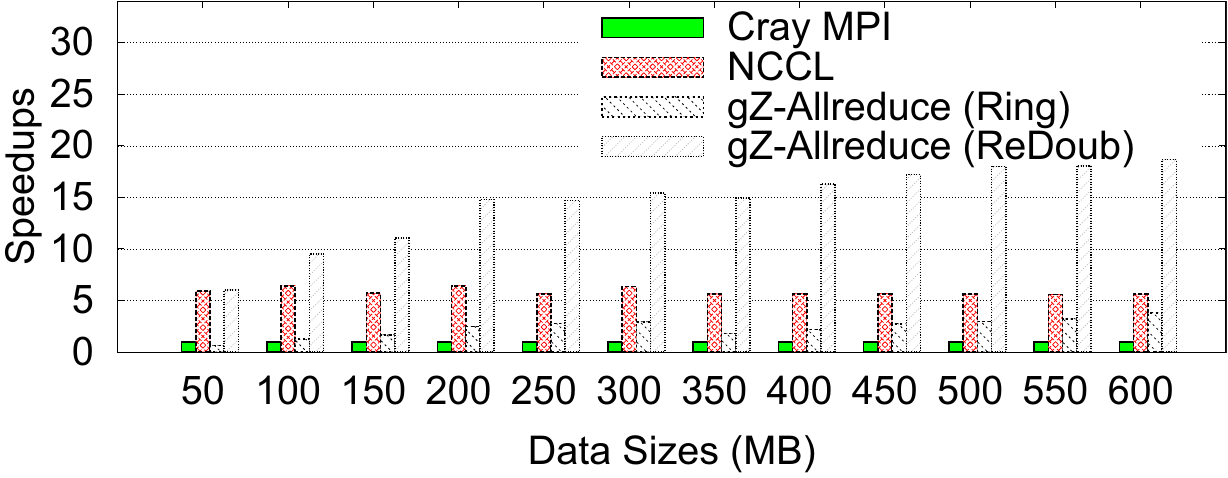}}
    \caption{Performance evaluation of our gZ-Allreduce with Cray MPI and NCCL in different data sizes.} 
    \label{fig-e2e-sizes-computation}
\end{figure}

\vspace{1mm}
\noindent\textbf{\textit{Evaluation with different message sizes.}} 
We evaluate the performance of our gZ-Allreduce algorithm using various data sizes up to 600 MB on a configuration of 64 NVIDIA A100 GPUs across 16 nodes. As observed in Figure \ref{fig-e2e-sizes-computation}, our recursive doubling-based gZ-Allreduce (ReDoub) consistently outperforms across all data sizes, achieving up to a speedup of 18.7$\times$ compared to Cray MPI and a 3.4$\times$ performance improvement over NCCL. Furthermore, with increasing data sizes, the speedup generally rises, demonstrating high scalability with respect to data size. The performance improvement originates from the significantly reduced message size and compression-related overheads in our \pname\ design, which can further mitigate network congestion with enlarging message sizes. However, the ring-based gZ-Allreduce (Ring), despite surpassing Cray MPI for the data size with 50+ MB, fails to outpace NCCL. This is attributed to the inefficient GPU utilization in gZ-Allreduce (Ring), which incurs substantial compression-related costs, outweighing the benefits of reduced message size.

\vspace{1mm}
\noindent\textbf{\textit{Evaluation with different GPU counts.}} 
In this section, we assess the scalability of our gZ-Allreduce algorithm with the complete RTM dataset of 646 MB data size, utilizing up to 512 NVIDIA A100 GPUs across 128 nodes. We start from 8 GPUs, as it is the minimal amount to have both internode and intranode communication with 4 GPUs per node. 

\begin{figure}[ht]
    \centering
    \hspace*{-5mm}
    {\includegraphics[scale=0.38]{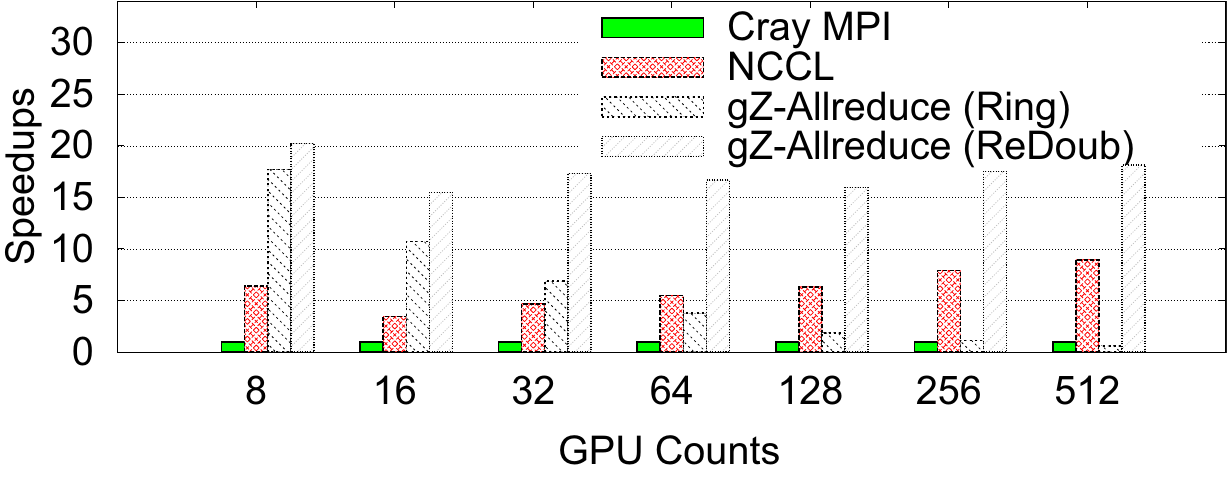}}
    \caption{Scalability evaluation of our gZ-Allreduce with Cray MPI and NCCL in different GPU counts.} 
    \label{fig-e2e-scales-computation}
\end{figure}

As depicted in Figure \ref{fig-e2e-scales-computation}, our recursive doubling-based gZ-Allred\linebreak uce (ReDoub) consistently performs the best, achieving up to 20.2$\times$ and 4.5$\times$ speedups compared to Cray MPI and NCCL respectively, across varying GPU counts. This superior performance stems from the substantial reduction in message size with relatively low compression cost achieved by our \pname\ framework.
When the GPU count is at 8, Cray MPI appears to suffer from significant performance degradation, as compared to the other three counterparts. Apart from the 8-GPU case, as the number of GPUs increases, both gZ-Allreduce (ReDoub) and NCCL tend to exhibit a greater performance boost compared to Cray MPI, indicating robust scalability with respect to the GPU count. This is because both gZ-Allreduce (ReDoub) and NCCL are optimized for large GPU count scenarios.
However, the trend differs for the ring-based gZ-Allreduce (Ring), which outperforms NCCL when the GPU count is 32 or less. As the GPU count increases, its performance deteriorates, ending up with the worst performance compared with other solutions in the case of 512 GPUs. The declining performance is attributed to the reduced input data size for each compression/decompression with an increase of GPU count, leading to lower device utilization and prolonged runtime, thus subpar scalability.

\subsubsection{Collective data movement}
In this section, we assess the performance of our \pname\ collective data movement framework using the widely-used Scatter operation, comparing it with Cray MPI. We exclude NCCL from this comparison as it has no implementation for Scatter.

\begin{figure}[ht]
    \centering
    \hspace*{-5mm}
    {\includegraphics[scale=0.38]{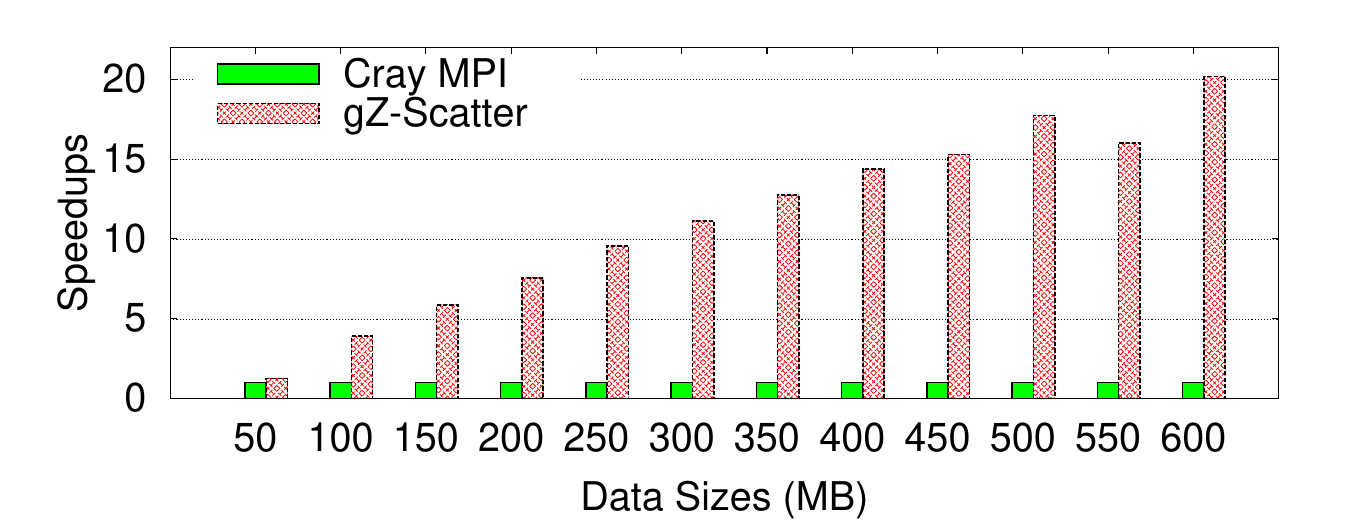}}
    \caption{Performance evaluation of our gZ-Scatter with Cray MPI in different data sizes.} 
    \label{fig-e2e-sizes-movement}
\end{figure}

\vspace{1mm}
\noindent\textbf{\textit{Evaluation with different message sizes.}} 
We evaluate the performance of our gZ-Scatter with data sizes up to 600 MB, using 64 NVIDIA A100 GPUs on 16 nodes. Figure \ref{fig-e2e-sizes-movement} indicates that our gZ-Scatter is able to consistently outperform Cray MPI across all data sizes. The speedup of gZ-Scatter enhances as the data size increases, achieving its maximum (20.2$\times$) at 600 MB. This demonstrates superior scalability with respect to data sizes, which can be attributed to the reduced message sizes and overlapping of compression, kernel launching, and data movement in our \pname\ framework.

\begin{figure}[ht]
    \centering
    \hspace*{-5mm}
    {\includegraphics[scale=0.38]{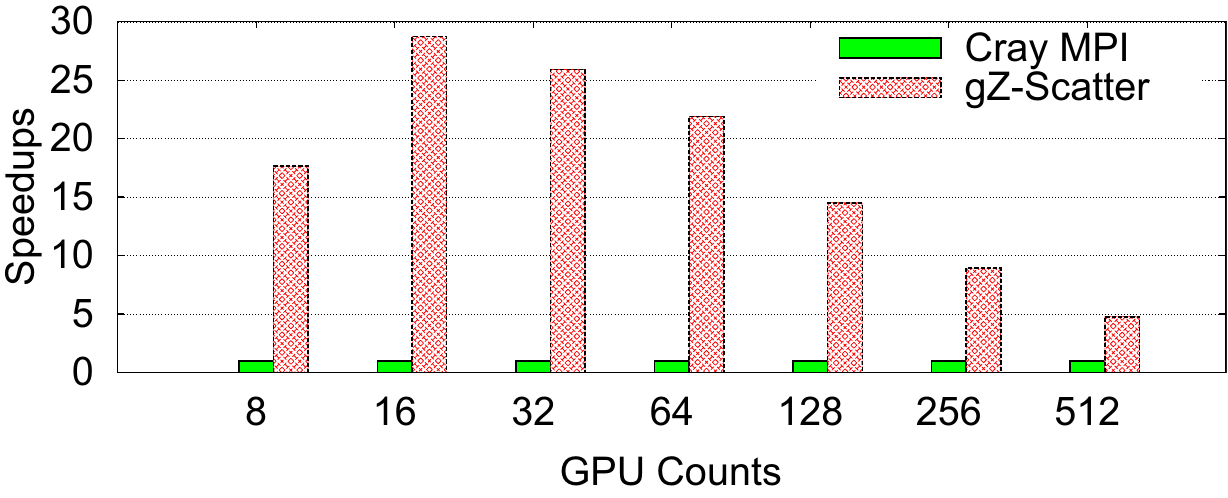}}
    \caption{Scalability evaluation of our gZ-Scatter with Cray MPI in different GPU counts.} 
    \label{fig-e2e-scales-movement}
\end{figure}

\vspace{1mm}
\noindent\textbf{\textit{Evaluation with different GPU counts.}} 
We assess the scalability of our gZ-Scatter with the complete RTM dataset, with a data size of 646 MB, using up to 512 NVIDIA A100 GPUs spread across 128 nodes. From Figure \ref{fig-e2e-scales-movement}, it is evident that our gZ-Scatter outperforms Cray MPI in all cases. As the GPU count increases, the speedup of gZ-Scatter first increases, peaking at 28.7$\times$, and then gradually decreases to 4.75$\times$ when the GPU count reaches 512.
Unlike the Allreduce scenario, the message size distributed to each non-root GPU in the Scatter communication pattern linearly decreases as the GPU count rises. When the GPU count is less than or equal to 16, the message size on the non-root GPU allows for high GPU utilization, hence the speedup grows with the increasing GPU count. However, when the GPU count exceeds or equals 32, the GPU utilization continues to drop, thereby reducing the collective performance and leading to a decrease in performance enhancement.

\subsection{Image Stacking Performance Evaluation with Accuracy Analysis}
\label{sec:image_stacking}

In this section, we employ the image stacking application to evaluate both the performance and accuracy of our {\pname}. Image stacking, a technique widely used in various scientific fields such as atmospheric science and geology, is employed to generate high-quality images by stacking multiple individual images, which essentially constitutes an Allreduce operation. As demonstrated by Gurhem in 2021 \cite{Gurhem2021Kirchhoff}, researchers use MPI to merge these individual images into a comprehensive final image.
As can be seen from Table \ref{tab:image_stacking_performance}, our ring-based {\pname} (Ring) outperforms Cray MPI by a factor of 3.99$\times$ when using an absolute error bound of 1E-4. Moreover, our recursive doubling-based {\pname} (ReDoub) offers even higher performance with speedups of up to 9.26$\times$ and 1.69$\times$ compared with Cray MPI and NCCL, respectively. This significant performance enhancement arises from the markedly reduced message sizes and compression-related overheads brought by our \pname\ framework.

The following text presents a performance breakdown analysis. For {\pname} (Ring), 84.08\% of the total runtime is consumed by compression, whereas {\pname} (ReDoub) has comparable compression and communication costs at 42.61\% and 46.28\% respectively. This substantial reduction in compression cost is due to higher GPU utilization and fewer compression operations in our optimized gZ-Allreduce (ReDoub) algorithm compared with {\pname} (Ring).

\begin{table}[ht]
\centering
\caption{Image stacking performance analysis (The speedups are based on Cray MPI. The last four columns are performance breakdowns of our {\pname}).}
\label{tab:image_stacking_performance}
\resizebox{1\columnwidth}{!}{%
\begin{tabular}{|c|c|cccc|}
\hline
 & \textbf{Speedups} & \multicolumn{1}{c|}{\textbf{Cmpr.}} & \multicolumn{1}{c|}{\textbf{Comm.}} & \multicolumn{1}{c|}{\textbf{Redu.}} & \textbf{Others} \\ \hline
\textbf{{\pname} (Ring)} & 3.99 & \multicolumn{1}{c|}{84.08\%} & \multicolumn{1}{c|}{14.08\%} & \multicolumn{1}{c|}{1.26\%} & 0.58\% \\ \hline
\textbf{{\pname} (ReDoub)} & 9.26 & \multicolumn{1}{c|}{42.61\%} & \multicolumn{1}{c|}{46.28\%} & \multicolumn{1}{c|}{11.04\%} & 0.06\% \\ \hline \hline
\textbf{NCCL} & 5.47 & \multicolumn{4}{c|}{No breakdown because of complexity} \\ \hline
\end{tabular}%
}
\end{table}

\begin{figure}[ht]
    \centering
    \subfloat[Cray MPI/NCCL (lossless)]{
        \includegraphics[scale=0.35]{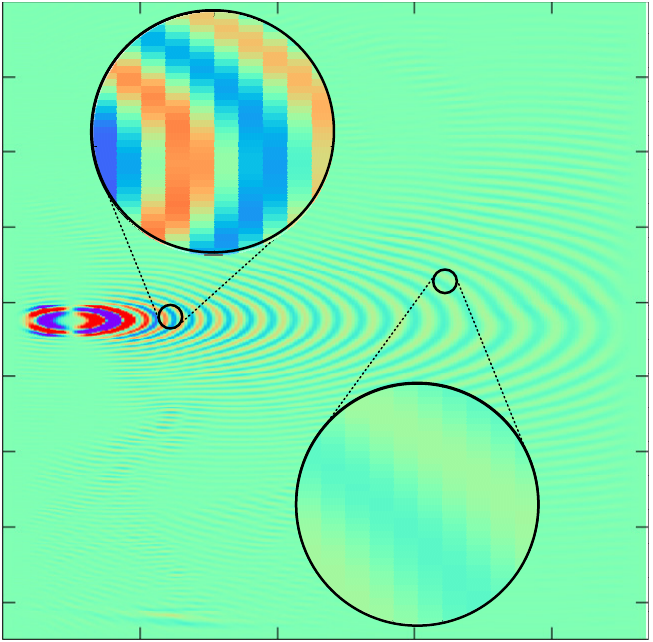}
        \label{fig-image-stacking-ori}
    }
    \subfloat[{\pname} (2E-4)]{
        \includegraphics[scale=0.35]{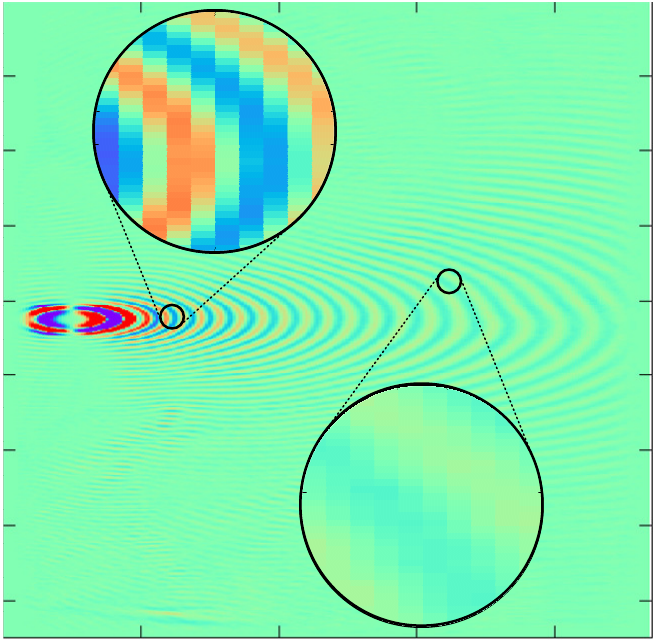}
        \label{fig-image-stacking-lossy}
    }
    \caption{Visualization of final stacking image.}
    \label{fig-image-stacking}
\end{figure}

In addition to performance analysis, we thoroughly evaluate the accuracy using both visualization method and numerical metrics such as the widely-used peak signal-to-noise ratio (PSNR) \cite{PSNR} and normalized root mean squared error (NRMSE) \cite{shcherbakov2013errormetrics}. Our accuracy-aware design allows {\pname} (ReDoub) to achieve excellent reconstructed image quality, even with an error bound of 2E-4, as shown in Figure \ref{fig-image-stacking}. The reconstructed image of {\pname} (Ring) also exhibits high visual quality, similar to that shown in Figure \ref{fig-image-stacking-lossy}, hence it is not presented separately here. When the error bound is tightened to 1E-4, as used in our performance analysis, {\pname} (Ring) reaches a great PSNR of 56.83 and an NRMSE of 1E-3. Meanwhile, {\pname} (ReDoub) demonstrates better data quality, achieving a PSNR of 57.80 and an NRMSE of 1E-3. The high accuracy of {\pname} confirms a controllable error propagation, which matches the theoretical analysis in \cite{huang2023ccoll}. {\pname} (ReDoub) exhibits a higher quality of reconstructed data over {\pname} (Ring), because of fewer error propagation steps as mentioned in Section \ref{sec:gZCCL-algorithm-design}.

\section{Conclusion and Future Work}
\label{sec:conclusion}
This paper presents \pname, an innovative framework that optimizes GPU-aware collective communications, offering minimized compression-related overheads and controlled accuracy. We devise two algorithm design frameworks and two collective optimization frameworks for both compression-enabled collective computation and collective data movement. We integrate the framework into a variety of collective communications including Allgather, Reduce\_scatter, Allreduce, and Scatter, demonstrating its generality. Our experiments with up to 512 NVIDIA A100 GPUs illustrate that our gZ-Allreduce surpasses the Allreduce operation in Cray MPI and NCCL by up to 20.2$\times$ and 4.5$\times$ respectively. In addition, our gZ-Scatter outperforms the Scatter operation in Cray MPI by 28.7$\times$, while NCCL lacks a Scatter implementation. In a nutshell, our work not only addresses the concerns of previous related efforts, such as inefficient GPU utilization, significant compression-related overheads, and inferior performance but also provides a groundwork for further studies in this domain.
Our future work will evaluate our {\pname} framework with more collective operations and we plan to extend {\pname} to more hardware such as FPGAs and AI accelerators.

\begin{acks}
This research was supported by the Exascale Computing Project (ECP), Project Number: 17-SC-20-SC, a collaborative effort of two DOE organizations -- the Office of Science and the National Nuclear Security Administration, responsible for the planning and preparation of a capable exascale ecosystem, including software, applications, hardware, advanced system engineering and early testbed platforms, to support the nation’s exascale computing imperative. The material was supported by the U.S. Department of Energy, Office of Science, Advanced Scientific Computing Research (ASCR), under Contract DE-AC02-06CH11357, and supported by the National Science Foundation under Grant OAC-2003709, OAC-2104023, and OAC-2311875. This research used resources from the Argonne Leadership Computing Facility, a U.S. DOE Office of Science user facility at Argonne National Laboratory, which is supported by the Office of Science of the U.S. DOE under Contract No. DE-AC02-06CH11357. 
\end{acks}

\end{document}